\documentclass[journal]{IEEEtran}

\usepackage{xcolor,soul,framed} 

\colorlet{shadecolor}{yellow}
\usepackage[pdftex]{graphicx}
\graphicspath{{../pdf/}{../jpeg/}}
\DeclareGraphicsExtensions{.pdf,.jpeg,.png}

\usepackage[cmex10]{amsmath}
\usepackage{array}
\usepackage{mdwmath}
\usepackage{mdwtab}
\usepackage{eqparbox}
\usepackage{url}

\begin{document}
\bstctlcite{IEEEexample:BSTcontrol}
    \title{TCAD model for TeraFET detectors operating in a large dynamic range}
  \author{Xueqing~Liu
      and~Michael~S.~Shur,~\IEEEmembership{Life Fellow,~IEEE}

  \thanks{(Corresponding author: Michael S. Shur.)}
  \thanks{Xueqing Liu and Michael S. Shur are with the Department of Electrical, Computer and Systems Engineering, Rensselaer Polytechnic Institute, 110 8th Street, Troy, NY 12180, USA (e-mail: liux29@rpi.edu, shurm@rpi.edu).}
}


\maketitle

\begin{abstract}
We present technology computer-aided design (TCAD) models for AlGaAs/InGaAs and AlGaN/GaN and silicon TeraFETs, plasmonic field effect transistors (FETs), for terahertz (THz) detection validated over a wide dynamic range. The modeling results are in good agreement with the experimental data for the AlGaAs/InGaAs heterostructure FETs (HFETs) and, to the low end of the dynamic range, with the analytical theory of the TeraFET detectors. The models incorporate the response saturation effect at high intensities of the THz radiation observed in experiments and reveal the physics of the response saturation associated with different mechanisms for different material systems. These mechanisms include the gate leakage, the velocity saturation and the avalanche effect.
\end{abstract}


%
\IEEEpeerreviewmaketitle


\section{Introduction}

\IEEEPARstart{T}{he} TeraFETs, plasmonic field effect transistors (FETs) applied in the terahertz (THz) frequency range, have been proposed since the early 1990s \cite{Dyakonov199310,Dyakonov199605,Dyakonov199610}. Over decades they have found wide applications in THz mixers \cite{Marinchio201001,Preu201205}, frequency multipliers \cite{Wu201801,Muralidharan2016}, transceivers \cite{Muralidharan2015,Wu201810}, imagers and sensors \cite{Knap200912,Lisauskas201401,Dyakonova201805}, etc. Recent interests include operating TeraFET detectors at high incident power, which could be used to measure the duration and structure of the high-power THz pulses \cite{Preu201511}. At large intensities of the incident THz radiation, the response has been observed to saturate in the measurements \cite{But2013}. Different propositions have been made to explain this effect \cite{But201404,Dyakonova201610,Dyakonova201509}. In this work we explore the reason for the response saturation effect by examining different physical mechanisms in the physics-based simulations using the validated TCAD models \cite{Liu2019}. The results reveal the physics of the response saturation associated with different mechanisms for different material systems. These mechanisms include the leakage current, the velocity saturation, and the avalanche effect. This insight into the device physics allows for the development of the next generation compact models for the TeraFET detectors that are valid over a wide dynamic operation range.

\begin{figure}
  \begin{center}
  \includegraphics[width=\columnwidth]{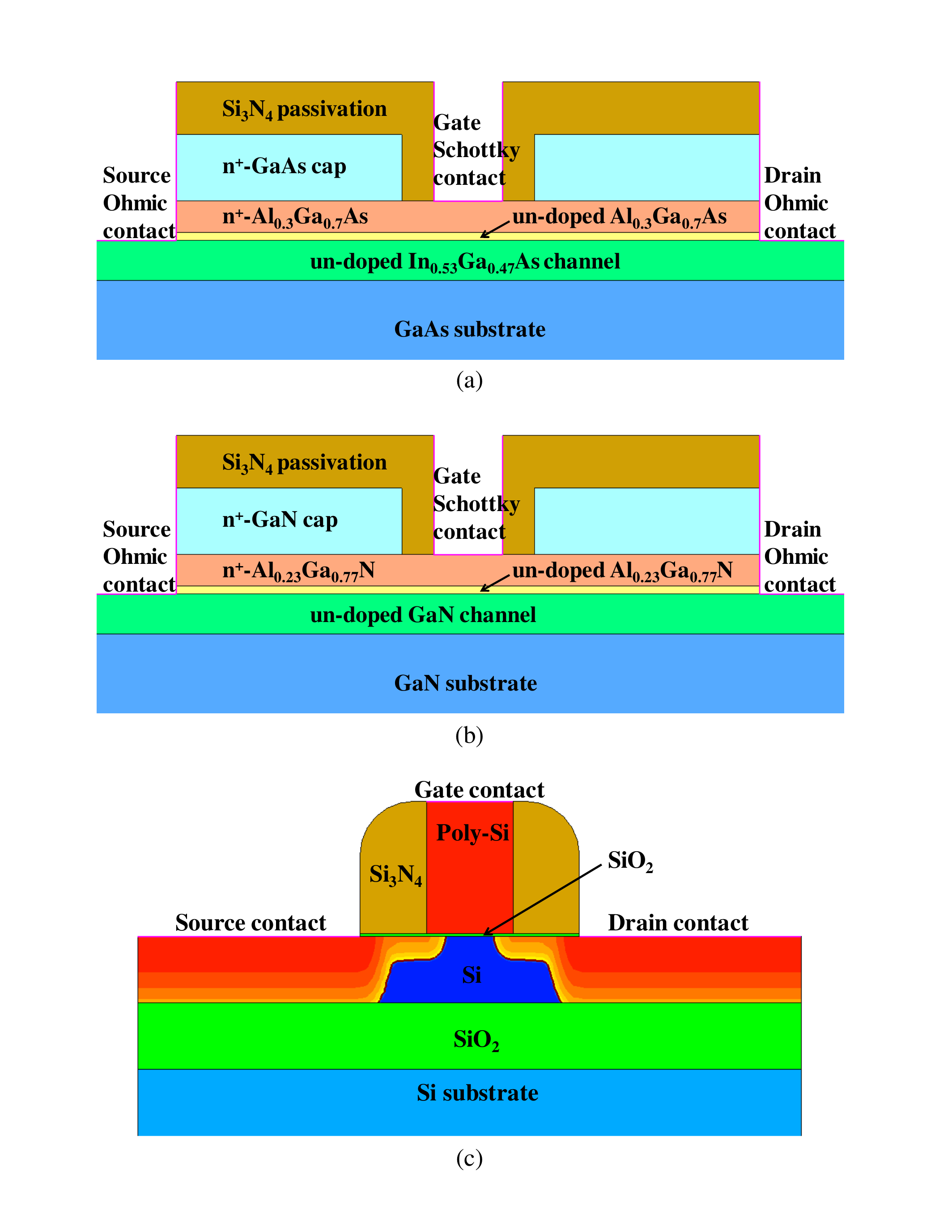}\\
  \caption{Schematic of the TeraFET structures in TCAD: (a) AlGaAs/InGaAs HFET, (b) AlGaN/GaN HFET, and (c) SOI MOSFET.}\label{fig1}
  \end{center}
\end{figure}

\begin{figure}[t]
  \begin{center}
  \includegraphics[width=\columnwidth]{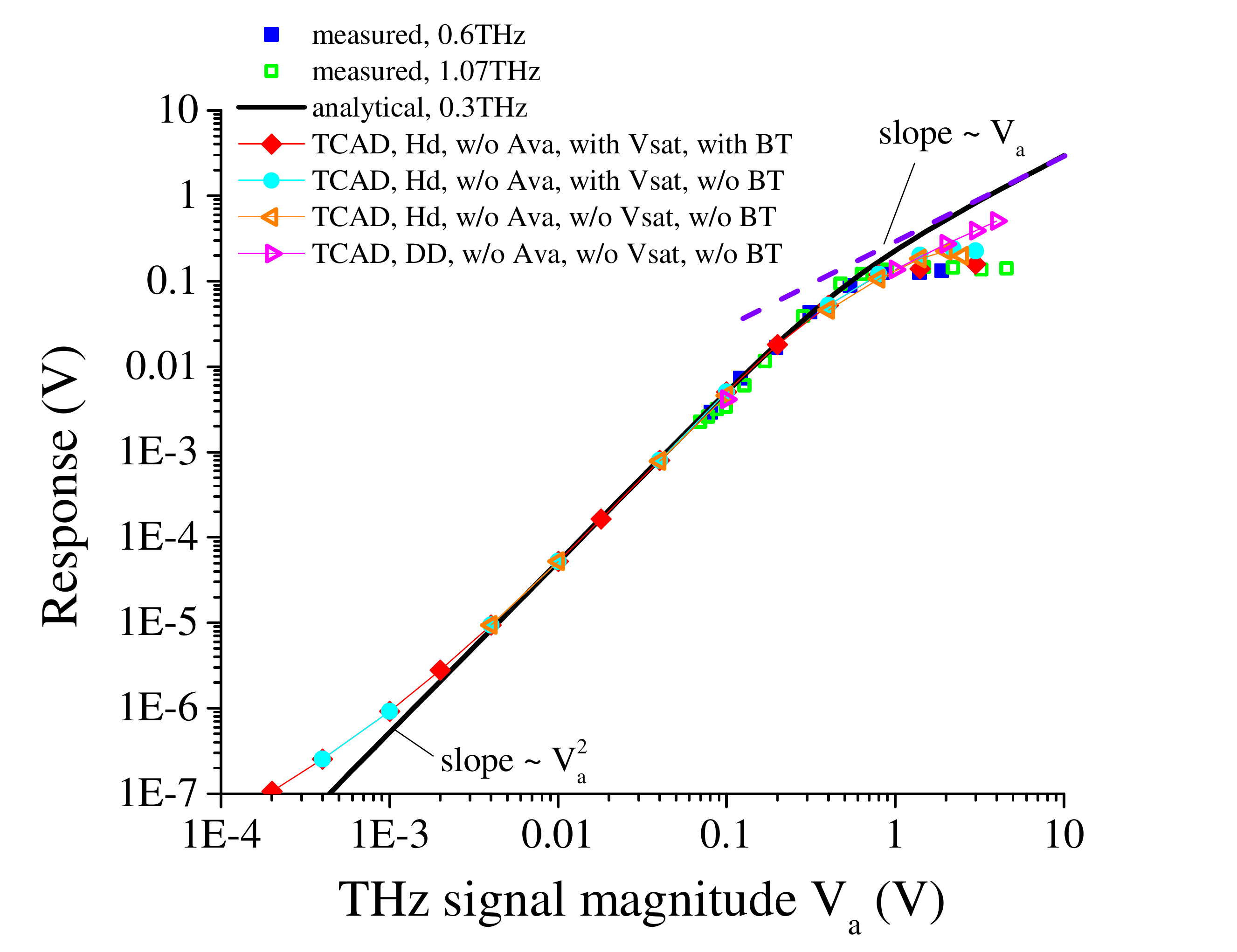}\\
  \caption{Comparison of analytical and simulated drain response at 0.3\,THz above threshold with the measured data in \cite{But2013} as a function of the THz signal magnitude for the AlGaAs/InGaAs HFET. The analytical and measured data are normalized to the range of the simulated results.}\label{fig2}
  \end{center}
\end{figure}

\begin{figure}[!h]
  \begin{center}
  \includegraphics[width=\columnwidth]{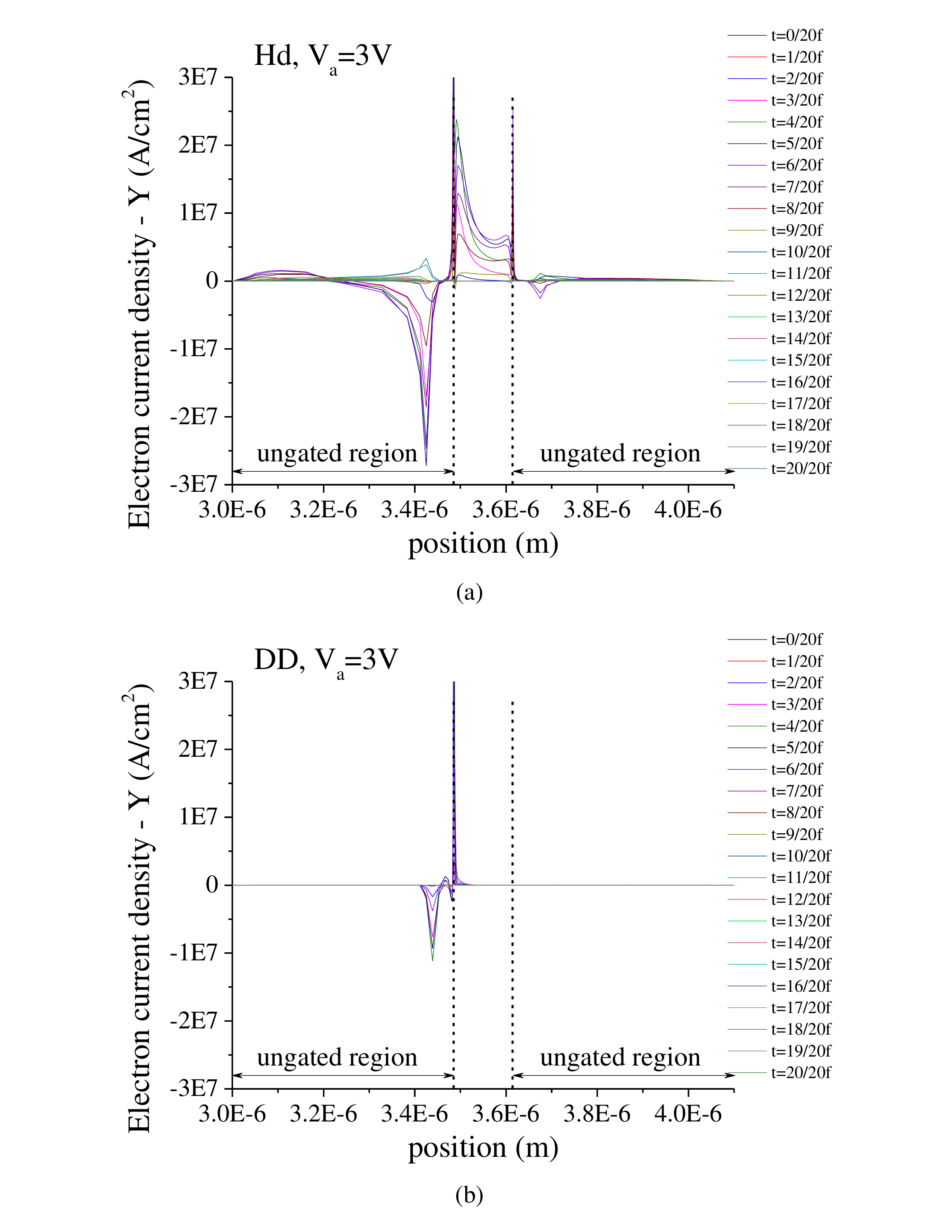}\\
  \caption{Profiles of the electron current density below the gate contact within a period for the AlGaAs/InGaAs HFET TCAD model at the high intensity level ($V_a=$ 3\,V) (a) with the hydrodynamic transport and (b) with the drift-diffusion transport. For both cases the avalanche, velocity saturation and barrier tunneling models are not included.}\label{fig3}
  \end{center}
\end{figure}

\begin{figure}[b]
  \begin{center}
  \includegraphics[width=\columnwidth]{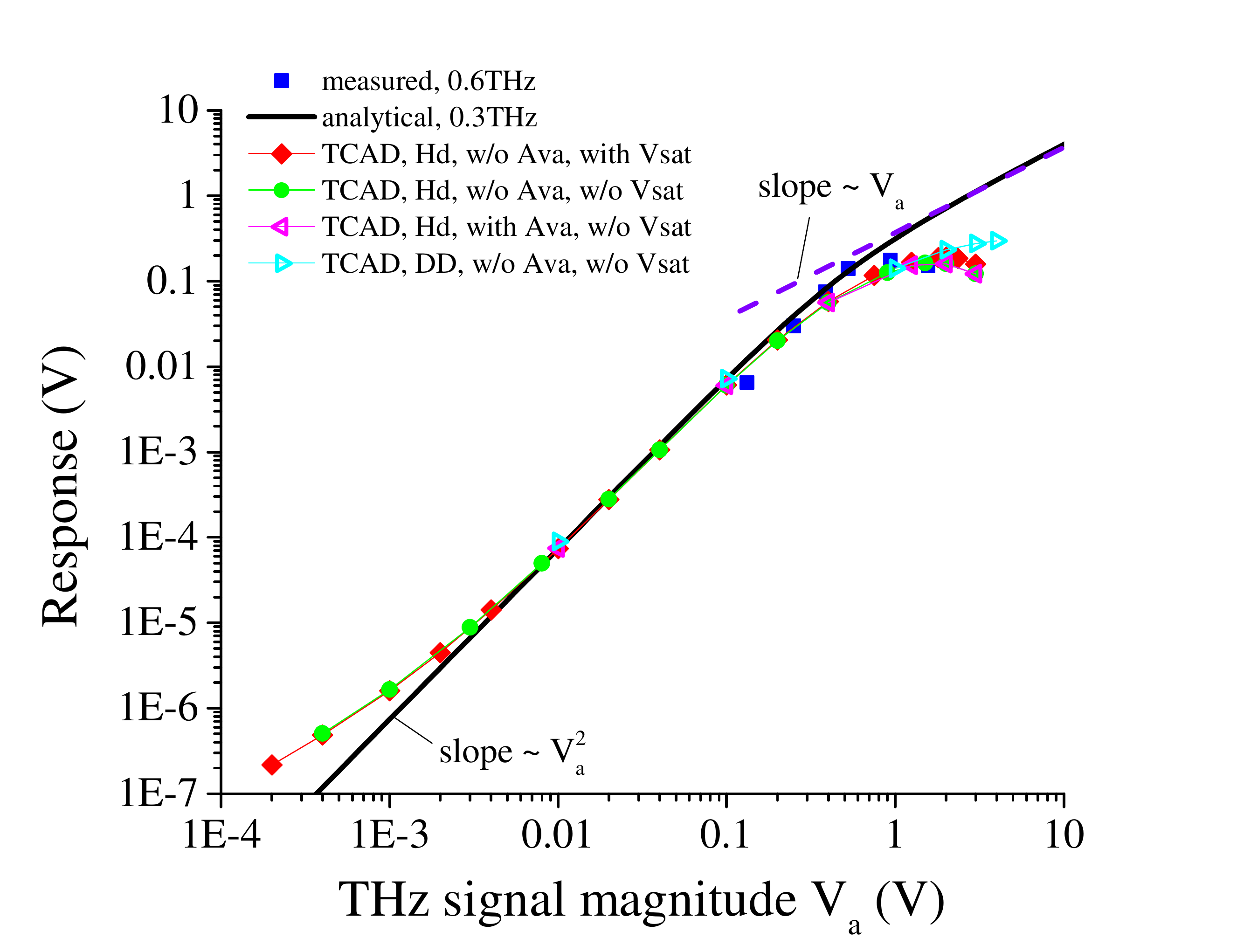}\\
  \caption{Comparison of analytical and simulated drain response at 0.3\,THz above threshold with the measured data in \cite{But2013} as a function of the THz signal magnitude for the AlGaN/GaN HFET. The analytical and measured data are normalized to the range of the simulated results.}\label{fig4}
  \end{center}
\end{figure}

\begin{figure}[t]
  \begin{center}
  \includegraphics[width=\columnwidth]{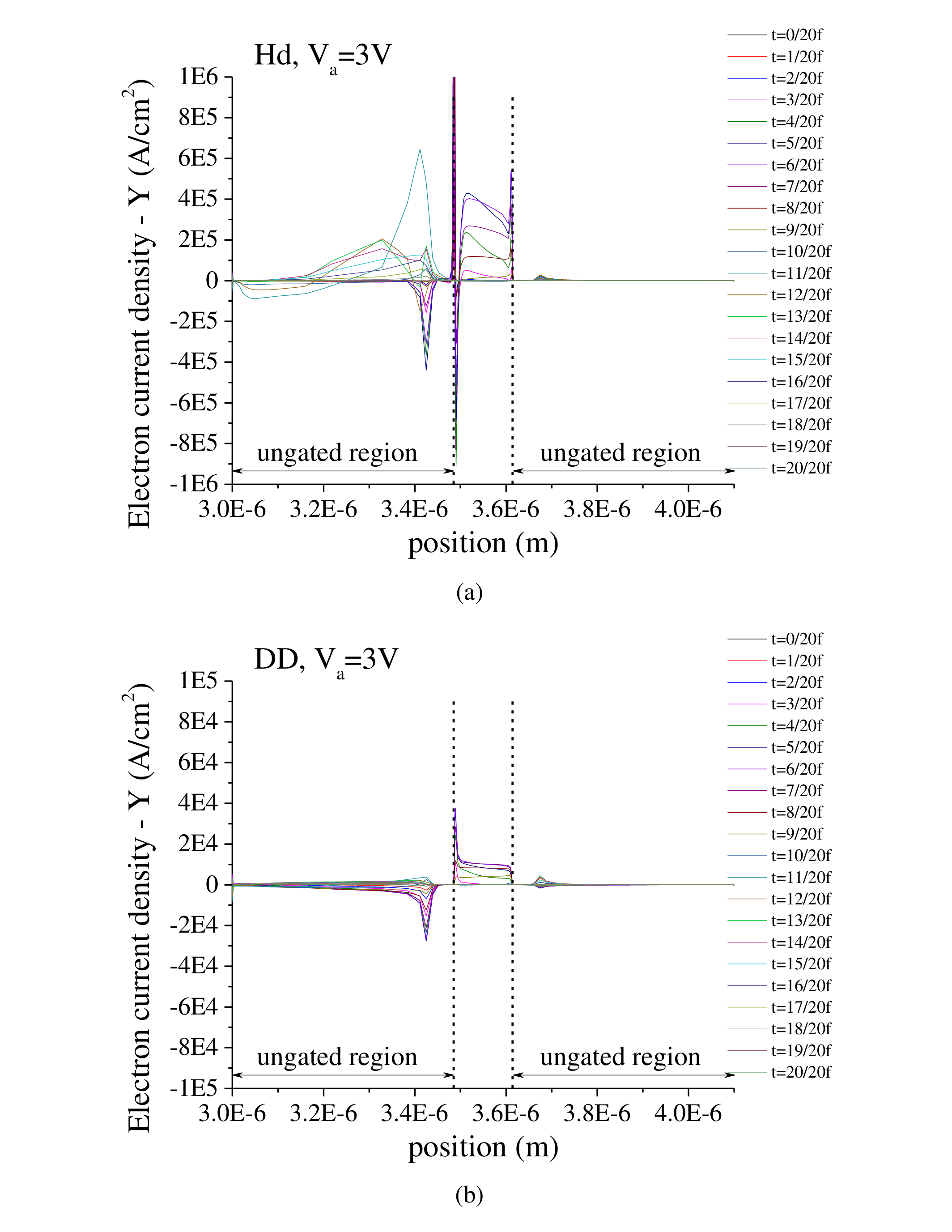}\\
  \caption{Profiles of the electron current density below the gate contact within a period for the AlGaN/GaN HFET TCAD model at the high intensity level ($V_a=$ 3\,V) (a) with the hydrodynamic transport and (b) with the drift-diffusion transport. For both cases the avalanche, velocity saturation and barrier tunneling models are not included.}\label{fig5}
  \end{center}
\end{figure}

\begin{figure}[t]
  \begin{center}
  \includegraphics[width=\columnwidth]{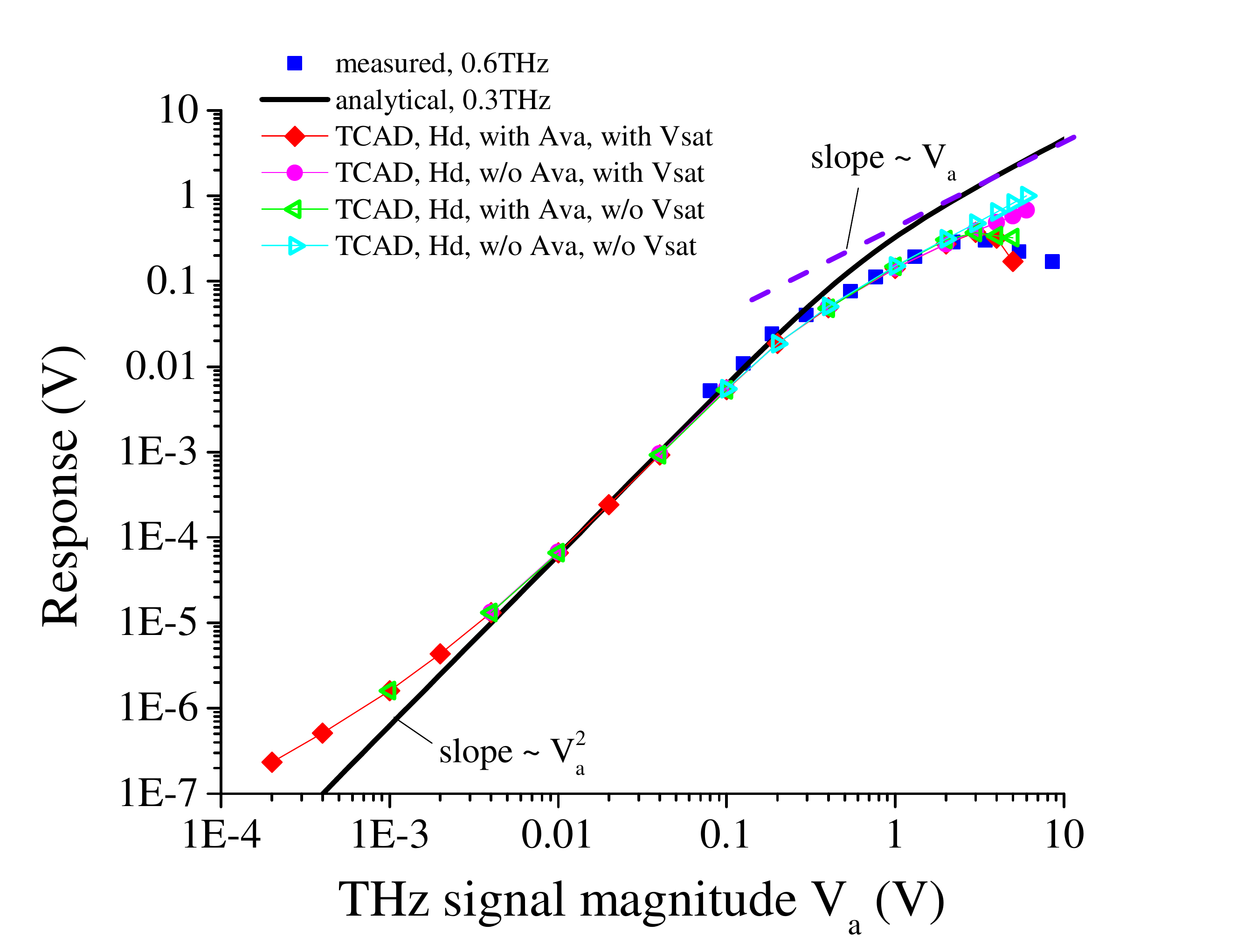}\\
  \caption{Comparison of analytical and simulated drain response at 0.3\,THz above threshold with the measured data in \cite{But2013} as a function of the THz signal magnitude for the SOI MOSFET. The analytical and measured data are normalized to the range of the simulated results.}\label{fig6}
  \end{center}
\end{figure}

\begin{figure}[!h]
  \begin{center}
  \includegraphics[width=\columnwidth]{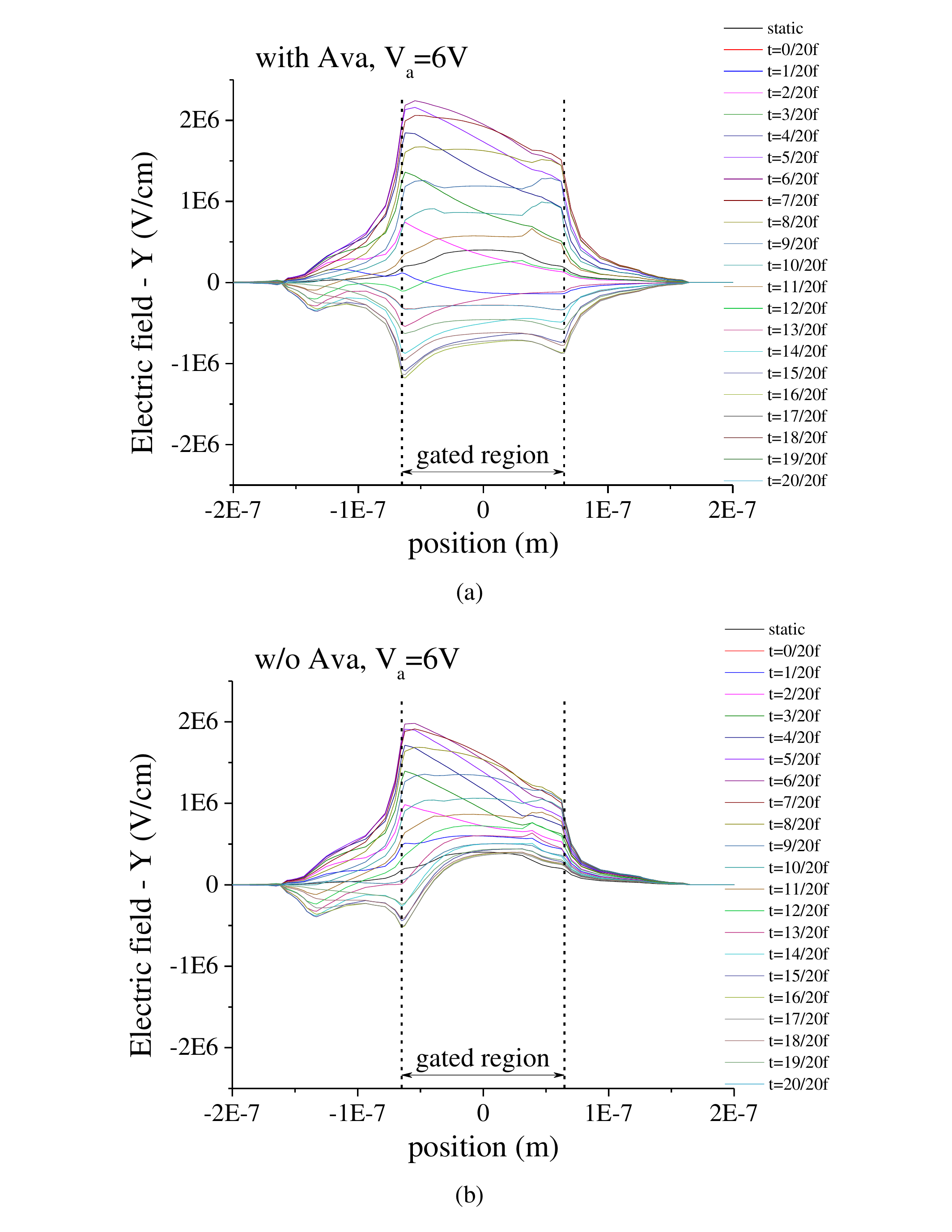}\\
  \caption{Profiles of the electric field along the channel within a period for the SOI MOSFET TCAD model at the high intensity level ($V_a=$ 6\,V) (a) with the avalanche model and (b) without the avalanche model. For both cases the hydrodynamic transport and the velocity saturation model are included.}\label{fig7}
  \end{center}
\end{figure}

\begin{figure}[h]
  \begin{center}
  \includegraphics[width=\columnwidth]{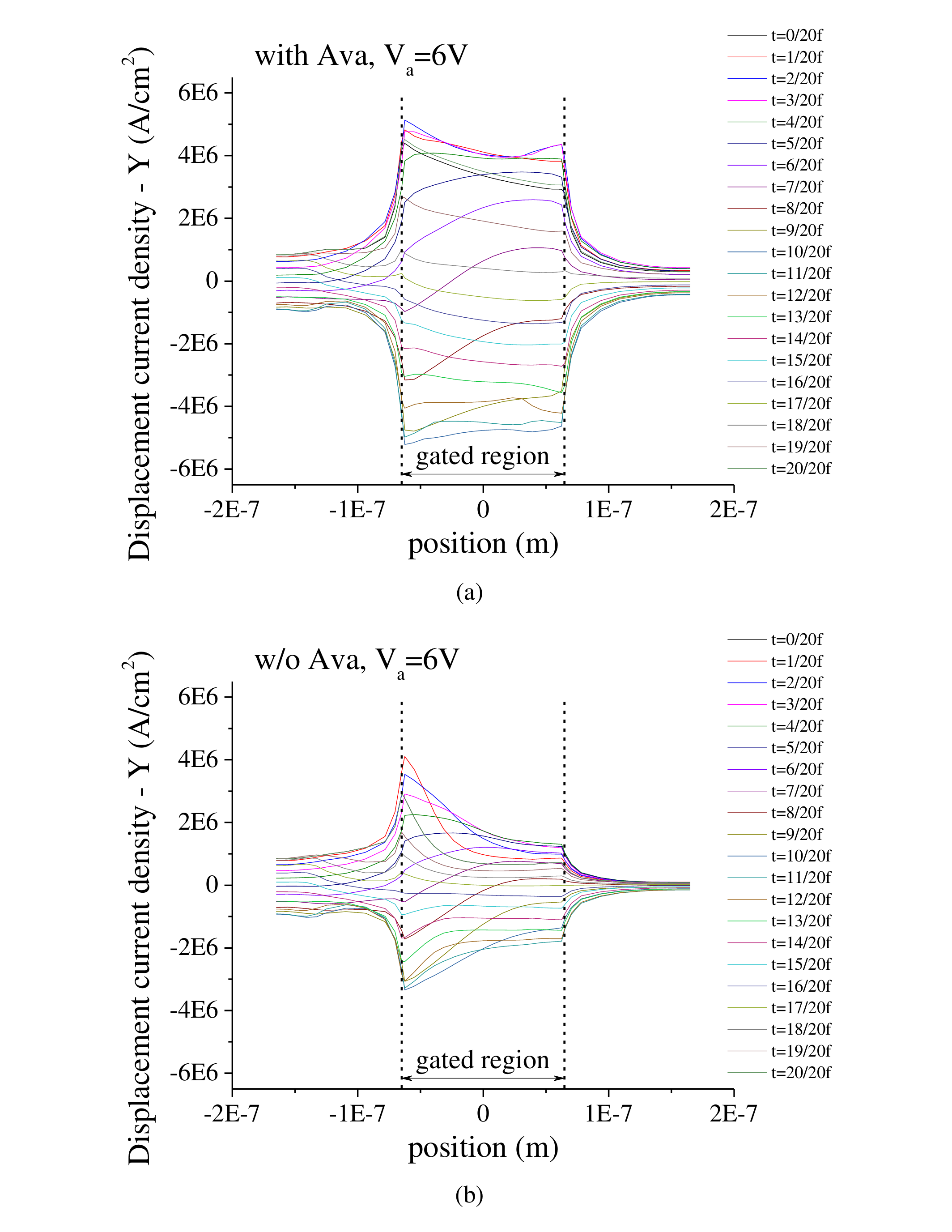}\\
  \caption{Profiles of the displacement current density below the top surface of the gate oxide within a period for the SOI MOSFET TCAD model at the high intensity level ($V_a=$ 6\,V) (a) with the avalanche model and (b) without the avalanche model. For both cases the hydrodynamic transport and the velocity saturation model are included.}\label{fig8}
  \end{center}
\end{figure}

\begin{figure}[h]
  \begin{center}
  \includegraphics[width=\columnwidth]{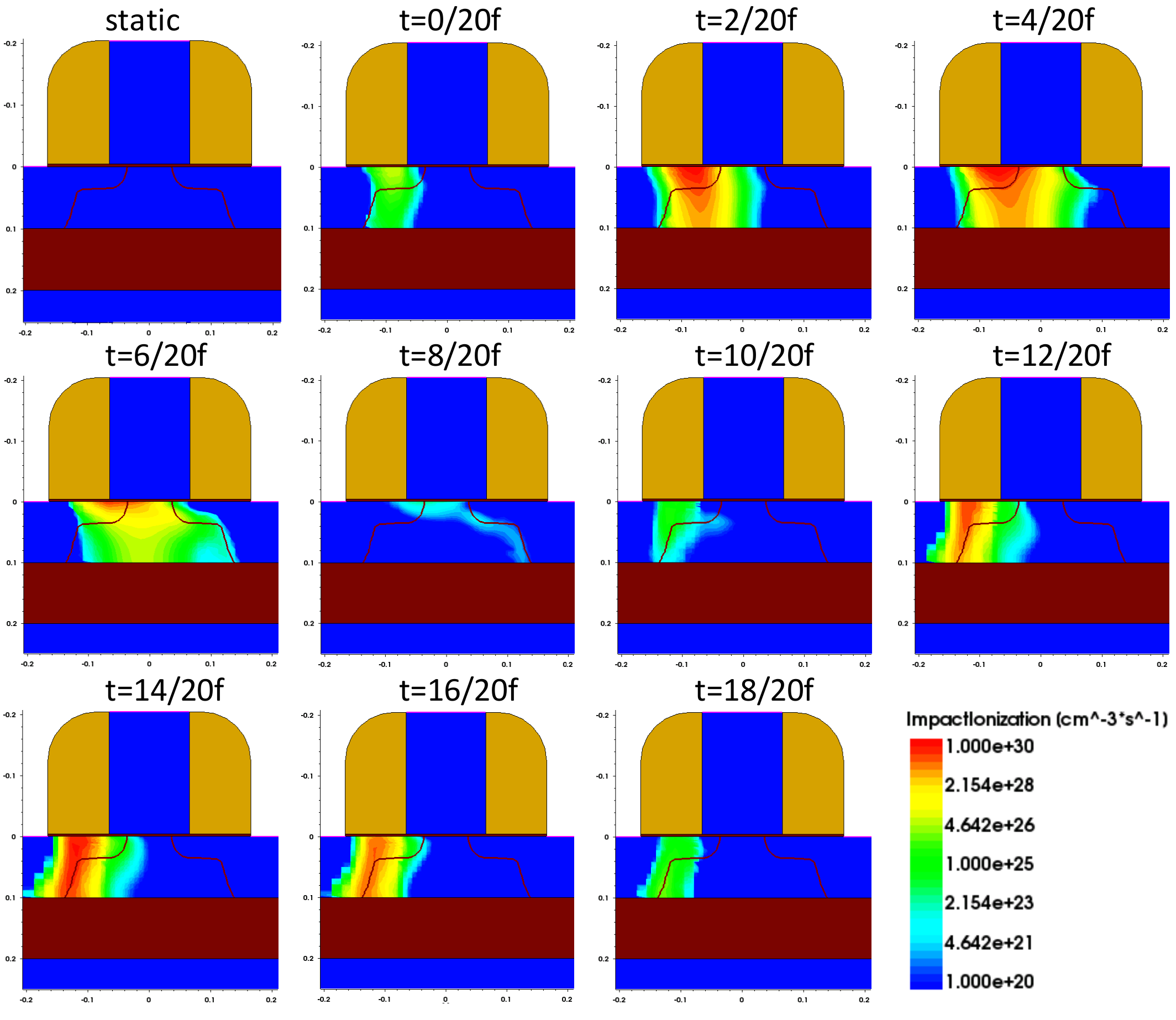}\\
  \caption{Profiles of the impact ionization generation rate at the high intensity level ($V_a=$ 6\,V) at different times within a period for the SOI MOSFET TCAD model with the avalanche effect.}\label{fig9}
  \end{center}
\end{figure}

\section{THz TCAD Models}

Fig.~\ref{fig1} shows the structures of the TeraFETs for different material systems in Synopsys Sentaurus TCAD. The TCAD models account for the hydrodynamic transport suitable for deep-submicron and heterostructure devices, and the velocity saturation, the generation-recombination, and the barrier tunneling mechanisms \cite{Sentaurus2017}. 

\subsection {AlGaAs/InGaAs HFETs}

The AlGaAs/InGaAs HFET TCAD model was built to simulate the 130\,nm AlGaAs/InGaAs pHEMTs fabricated by TriQuint Inc. (now Qorvo) \cite{Ping200504}. It has been validated by comparing the simulated I-V characteristics and the dependence of the THz response on the gate bias with the measured data and the analytical results \cite{Liu2019}.

Fig.~\ref{fig2} shows the dependence of the simulated detector response at 0.3\,THz above threshold ($V_{gt}=$ 0.12\,V) on the THz signal magnitude $V_a$ for the AlGaAs/InGaAs HFET TCAD model, compared with the measured data in \cite{But2013} and the analytical results with the open boundary condition at the drain. To explore the device physics for the response saturation, different physical mechanisms including the hydrodynamic (Hd) or drift-diffusion (DD) transport, avalanche (Ava), velocity saturation (Vsat), and gate barrier tunneling (BT) were turned on and off in the simulation. It could be seen that the simulated response gives a quadratic response (proportional to $V_a^2$) at low intensities, which is consistent with the analytical result from around 5\,mV to 1\,V. At high intensities ($V_a>$ 1 V), the response saturation observed in the measurements could also be demonstrated in the TCAD model, but not in the analytical result, since the analytical theory uses the next terms in the Taylor series expansion of the response dependence on the THz voltage \cite{Gutin201207}. Hence, the analytical theory applies to an intermediate range of powers but it cannot reproduce the entire dynamic range. Also, the model with the drift-diffusion transport does not show the response saturation effect at high intensities of the incident THz radiation ($V_a>$ 1\,V), while the models with the hydrodynamic transport show the response saturation behavior. The main difference between the hydrodynamic transport and the drift-diffusion transport in TCAD is that the hydrodynamic model accounts for the energy transport across heterointerfaces, which is not considered in the drift-diffusion model. Therefore, the hydrodynamic transport could lead to larger gate leakage than the drift-diffusion transport. Hence, the response saturation could be associated with the gate leakage which contributes to the rectification of the input THz signal at the gate. Fig.~\ref{fig3} shows the profiles of the electron current density transverse to the conducting channel direction at the positions closely (1\,nm) below the Schottky gate contact at different times within a period for $V_a=$ 3\,V using the hydrodynamic and drift-diffusion models. As seen, the profiles with the hydrodynamic transport show much larger gate leakage current than the drift-diffusion transport.

\subsection {AlGaN/GaN HFETs}

The TCAD model developed for the AlGaN/GaN HFET uses the same dimensions with the AlGaAs/InGaAs HFET TCAD model, as shown in Fig.~\ref{fig1} (b). Fig.~\ref{fig4} shows the simulated detector response at 0.3\,THz above threshold ($V_{gt}=$ 0.12\,V) as a function of the THz signal magnitude $V_a$ for the AlGaN/GaN HFET TCAD model with different physical mechanisms, compared with the measured data in \cite{But2013} and the analytical results with the open boundary condition at the drain. Again, from the profiles of the electron current density transverse to the channel at the positions closely (1\,nm) below the Schottky gate contact in Fig.~\ref{fig5}, the response saturation could be linked to the higher gate leakage current observed in the hydrodynamic transport than in the drift-diffusion transport. 

\subsection {Si MOSFETs}

The response saturation at high intensities of the THz radiation was also observed for Si MOSFETs \cite{But2013}. Since the gate leakage could be negligible in Si MOSFETs due to the gate insulator, other mechanisms could be responsible for the response saturation. To investigate this, the TCAD model in Fig.~\ref{fig1} (c) is considered. The model is set up based on an exemplary silicon-on-insulator (SOI) N-channel MOSFET using the default material parameter files for silicon in Sentaurus TCAD \cite{Sentaurus2017}. The gate length for the model is set as 130\,nm, the same as for the HFET models. The response saturation could also be demonstrated at large intensities with the SOI MOSFET TCAD model, as shown in Fig.~\ref{fig6}. The comparison of the simulated results with different mechanisms shows that the response saturation could be associated with the avalanche effect and also affected by the velocity saturation. The electric field in the gate oxide was checked to be below the SiO$_{2}$ breakdown field (10\,MV/cm) indicating no gate breakdown. However, at high THz fields, carriers could be generated from impact ionization and travel into the channel and change the electric field which could affect the response at the drain. This effect of the avalanche model could be seen in Fig.~\ref{fig7} which shows the profiles of the electric field transverse to the channel at the positions along the channel and closely (1\,nm) below the SiO$_{2}$/Si interface within a period above threshold ($V_{gt}=$ 0.2\,V). Also, although the gate conduction current is zero, the displacement current at the gate exists, as shown in Fig.~\ref{fig8}. The inclusion of the avalanche model leads to higher displacement current near the drain, which could play a similar role as the gate leakage in the HFETs. Fig.~\ref{fig9} shows the profiles of the impact ionization rate at $V_a=$ 6\,V at different times within a period for the SOI MOSFET TCAD model with the avalanche effect, indicating where the carriers could be generated at high intensities of the incident THz radiation.

\section{Conclusion}
In this work, the TCAD models valid in a large dynamic range were presented. The TCAD models explain the experimentally observed response saturation at high intensity levels of the incident THz radiation (above 1\,V). By activating or deactivating different physical mechanisms in the TCAD models, the reason for the response saturation effect was found out to be associated with the gate leakage for AlGaAs/InGaAs HFETs and AlGaN/GaN HFETs and affected by the velocity saturation and the avalanche effect for Si MOSFETs.



%





\ifCLASSOPTIONcaptionsoff
  \newpage
\fi

\vfill


\end{document}